\documentclass[prd,aps,floats,showpacs,twocolumn,twoside,preprintnumbers,
superscriptaddress,floatfix]{revtex4}

\usepackage{bm}
\usepackage[T1]{fontenc}
\usepackage[latin1]{inputenc}
\usepackage{amsmath}
\usepackage{amssymb}
\usepackage{mathrsfs}
\usepackage{epsfig}

\begin{document}


\title{Neutron star cooling constraints 
for color superconductivity in hybrid stars}

\author{S. Popov}
\email{polar@sai.msu.ru}
\affiliation{Sternberg Astronomical Institute,
Universitetski pr. 13, 119992 Moscow, Russia}
\author{H. Grigorian}
\email{hovik.grigorian@uni-rostock.de}
\affiliation{Institut f\"ur Physik, Universit\"at Rostock, 
D-18051 Rostock, Germany, and\\
Department of Physics, Yerevan State University, 375025 Yerevan, Armenia}
\author{D. Blaschke}
\email{Blaschke@theory.gsi.de}
\affiliation{Gesellschaft f\"ur Schwerionenforschung mbH (GSI),
D-64291 Darmstadt, Germany, and\\
Bogoliubov Laboratory for Theoretical Physics, JINR Dubna, 141980 Dubna, 
Russia}

\date{\today}

\begin{abstract}
We apply the recently developed
$\mathrm{Log\, N}$-$\mathrm{Log\, S}$ test 
of compact star cooling theories for the first time
to hybrid stars with a color superconducting quark matter core. 
While there is not yet a microscopically founded superconducting quark matter 
phase which would fulfill constraints from cooling phenomenology, we explore
the hypothetical 2SC+X phase 
and show that the magnitude and density-dependence of the X-gap can be chosen 
to satisfy a set of tests: temperature - age ($\mathrm{T}$-$\mathrm{t}$), the 
brightness constraint, $\mathrm{Log\, N}$-$\mathrm{Log\, S}$, and the mass 
spectrum constraint.
The latter test appears as a new conjecture from the present investigation. 
\end{abstract}


\maketitle

\section{Introduction}
\label{intro}

Recently, preparations for terrestrial laboratory experiments with heavy-ion 
collisons have been started, where it is planned to access the high-density/
low temperature region of the QCD phase diagram and explore physics at the 
phase boundary between hadronic and quark matter, e.g., within the CBM 
experiment at FAIR Darmstadt.

Predictions for critical parameters in  this domain of the temperature-density 
plane are uncertain since they cannot be checked against Lattice-QCD 
simulations which became rather precise at zero baryon densities.
Chiral quark models have been developed and calibrated with these results.
They can be extended into the finite-density domain and suggest a rich 
structure of color superconducting phases. 
These hypothetical phase structures shall imply consequences for the structure 
and evolution of compact stars, where the constraints from mass and radius 
measurements as well as from the cooling phenomenology have recently reached an
unprecedented level of precision which allows to develop decisive tests of
models for high-density QCD matter.

Among compact stars (we will address them also with the general term 
{\it neutron stars} (NSs)) one can distiguish three main classes according 
to their composition: 
hadron stars, quark stars (bare surfaces or with thin crusts), 
and hybrid stars (HyS). The latter are the subject of the present study.

Observations of the surface thermal emission of NSs 
is one of the most
promising ways to derive detailed information about processes in interiors
of compact objects (see \cite{Page:2005fq,Page:2004fy,Yakovlev:1999sk} 
for recent reviews).
In \cite{Popov:2004ey} (Paper I hereafter) we
proposed to use a population synthesis of close-by cooling NSs as
an addtional test for theoretical cooling curves. 
This tool, based on calculation of the $\mathrm{Log\, N}$-$\mathrm{Log\,S}$ 
distribution, was shown to be an effective supplement to the standard 
$\mathrm{T}$~-~$\mathrm{t}$ (Temperature vs. age) test. 
In Paper I we used cooling curves for hadron stars calculated
in \cite{Blaschke:2004vq}.
Here we study cooling curves of HyS
calculated in \cite{Grigorian:2004jq} (Paper II hereafter).

Except $\mathrm{T}$~-~$\mathrm{t}$ and $\mathrm{Log\, N}$-$\mathrm{Log\,S}$
we use also the brightness constraint test (BC) suggested in 
\cite{Grigorian:2005fd}.
We apply altogether three tests -- $\mathrm{T}$-$\mathrm{t}$,
$\mathrm{Log\, N}$-$\mathrm{Log\,S}$, and BC  -- to five sets of cooling
curves of HyS. In the next section we describe calculation of these curves.
In Section III we discuss the population synthesis scenario. 
After that we present our results which imply the conjecture of a new mass 
spectrum constraint from Vela-like objects.
In Section 5 we discuss the results and present our conclusions in Section ~6. 

\section{Cooling curves for hybrid stars}
\label{cool}


\subsection{Hybrid stars}

The description of compact star cooling with color superconducting
quark matter interior is based on the approach introduced in Paper II
which will be briefly reviewed here, see also \cite{Blaschke:2005dc} for 
a recent summary. 
A nonlocal, chiral quark model is employed which supports 
compact star configurations with a rather large quark core
due to the relatively low critical densities for the
deconfinement phase transition from hadronic matter to color superconducting 
quark matter. 
In the interior of the compact star in late cooling
stages, when the temperature is well below the opacity temperature
$T_{\rm opac}\sim 1$ MeV for neutrino untrapping, four phases of
quark matter are possible: normal quark matter (NQ),
two-flavor superconducting matter (2SC), a mixed phase of both
(NQ-2SC) and the color-flavor-locking phase (CFL). The
state-of-the-art calculations for a three-flavor quark matter
phase diagram within a chiral (NJL) quark model of quark matter
and selfconsistently determined quark masses are described in
Refs. \cite{Ruster:2005jc,Blaschke:2005uj,Abuki:2004zk}.

The detailed structure of the phase diagram in these models still
depends on the strength parameter $G_D$ of the diquark coupling
(and on the formfactor of the momentum space regularization, see
\cite{Aguilera:2004ag}). For all values of  $G_D$ no stable hybrid
stars with a CFL phase have been found yet, see
\cite{Buballa:2003qv}, and Refs. therein. 
We will restrict us here to the discussion of 2SC and NQ phases. 

The 2SC phase occurs at lower baryon densities than the CFL phase 
\cite{Steiner:2002gx,Neumann:2002jm}. 
For applications to compact stars the omission
of the strange quark flavor is justified by the fact that chemical
potentials in central parts of the stars barely reach the
threshold value at which the mass gap for strange quarks breaks
down and they may appear in the system \cite{Gocke:2001ri}.

It has been shown in \cite{Blaschke:2003yn} that a nonlocal chiral quark
model with the Gaussian formfactor ansatz leads to an early onset of the 
deconfinement transition so that hybrid stars with
large quark matter cores \cite{Grigorian:2003vi} can be discussed.

In describing the hadronic part of the hybrid star, as in
\cite{Blaschke:2004vq}, we adopt the Argonne $V18+\delta v+UIX^*$ model
for the EoS \cite{Akmal:1998cf}, which is based on 
recent data for the nucleon-nucleon interaction with the
inclusion of a parameterized three-body force and relativistic
boost corrections.

Actually we continue to adopt an analytic parameterization of this
model by Heiselberg and Hjorth-Jensen \cite{Heiselberg:1999fe}, where the fit
is done for $n<4~n_0$ with $n_0=0.16$ fm$^{-3}$ being the nuclear saturation 
density. This EoS  fits the symmetry energy  to the
original Argonne $V18+\delta v +UIX^*$ model in the mentioned
density interval and smoothly incorporates causality constraints at high 
densities. 
The threshold density for the DU process is $n_c^{\rm DU}\simeq~5.19~n_0$, 
i.e. it occurs in stars with masses exceeding 
$M_c^{\rm DU}\simeq 1.839~M_{\odot}$).

\subsection{Cooling}

For the calculation of the cooling of the hadronic part of the
hybrid star we use the same model as in \cite{Blaschke:2004vq}. 
The main processes are the medium modified Urca (MMU) and the pair breaking
and formation (PBF) processes for our adopted EoS of hadronic
matter.  
For a recent, more detailed discussion of these processes and the role
of the $3P_2$ gap, see \cite{Grigorian:2005fn}. 

The possibilities of pion condensation and of other so called
exotic processes are suppressed since in the model \cite{Blaschke:2004vq}
these processes may occur only for neutron star masses exceeding
$M_c^{\rm quark}= 1.214~M_{\odot}$. The DU process is irrelevant
in this model up to very large neutron star masses
$M>1.839~M_{\odot}$. The $1S_0$ neutron and proton gaps are taken
the same as those shown by thick lines in Fig. 5 of  Ref.
\cite{Blaschke:2004vq}. We pay particular attention to the fact that the
$3P_2$ neutron gap is additionally suppressed by the factor $0.1$
compared to that shown in Fig. 5 of \cite{Blaschke:2004vq}. This
suppression  is motivated by  the result of the recent work in
\cite{Schwenk:2003bc} and is required to fit the cooling data.

For the calculation of the cooling of the quark core in the hybrid
star we use the model introduced in \cite{Blaschke:2004vq}. 
We incorporate the most
efficient processes: the quark direct Urca (QDU) processes on
unpaired quarks, the quark modified Urca (QMU), the quark
bremsstrahlung (QB), the electron bremsstrahlung (EB), and the
massive gluon-photon decay (see \cite{Blaschke:1999qx}). 
Following \cite{Jaikumar:2001hq}
we include the emissivity of the quark pair formation and breaking
(QPFB) processes. The specific heat incorporates the quark
contribution, the electron contribution and the massles and
massive gluon-photon contributions. The heat conductivity contains
quark, electron and gluon terms.

The 2SC phase has one unpaired color of quarks (say blue) for which
the very effective quark DU process works and leads to a too fast
cooling of the hybrid star in disagreement with the data
\cite{Grigorian:2004jq}. We have suggested to assume a weak pairing channel
which could lead to a small residual pairing of the hitherto
unpaired blue quarks. We call the resulting gap $\Delta_X$ and
show that for a density dependent ansatz

\begin{equation}
\Delta_{\mathrm{X}}= \Delta_0 \, \exp{\left[-\alpha\, \left(
\frac{\mu - \mu_c}{\mu_c}\right)\right]} 
\label{gap}
\end{equation}

with $\mu$ being the quark chemical potential, $\mu_c=330$ MeV.
Here we use different values of $\alpha$ and $\Delta_0$, which  are 
given in Table 1 characterizing the model.

The physical origin of the X-gap remains to be identified. It
could occur, e.g., due to quantum fluctuations of color neutral
quark sextett complexes \cite{Barrois:1977xd}. Such calculations have not
yet been performed with the relativistic chiral quark models.

For sufficiently small $G_D$, the 2SC pairing may be inhibited at
all. In this case, due to the absence of this competing spin-0
phase with large gaps, one may invoke a spin-1 pairing channel in
order to avoid the DU problem. In particular the
color-spin-locking (CSL) phase \cite{Schafer:2000tw} may be in accordance
with cooling phenomenology as all quark species are paired and the
smallest gap channel may have a behavior similar to Eq.
(\ref{gap}), see \cite{Aguilera:2005tg}. A consistent cooling
calculation for this phase, however, requires the evaluation of
neutrino emissivities (see, e.g. \cite{Schmitt:2005wg} and references therein) 
and transport coefficients, which is still to be performed.

Gapless superconducting phases can occur when the diquark coupling
parameter is small so that the pairing gap is of the order of the
asymmetry in the chemical potentials of the quark species to be
paired. Interesting implications for the cooling of gapless CFL
quark matter have been conjectured due to the particular behavior
of the specific heat and neutrino emissivities \cite{Alford:2004zr}.
For reasonable values of $G_D$, however, these phases
do occur only at too high temperatures to be relevant for late
cooling, if a stable hybrid configuration with these phases could
be achieved at all \cite{Blaschke:2005uj}.

The weak pairing channels are characterized by gaps typically in
the interval $10~$ keV $\div 1~$MeV, see discussion of
different attractive interaction channels in paper \cite{Alford:2002rz}.


\begin{figure}
\includegraphics[width=0.46\textwidth,angle=-90]{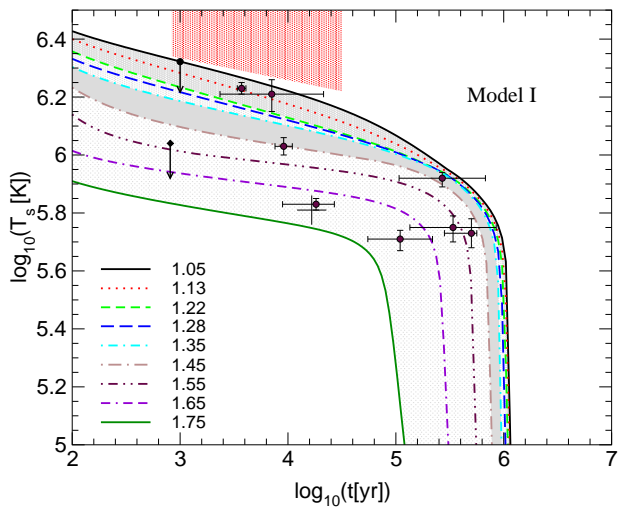}
\caption[]{Hybrid star cooling curves for Model I. 
Different lines correspond to compact star mass values indicated in the legend 
(in units of $M_\odot$), data points with error bars are taken from Ref. 
\cite{Page:2004fy}. For the explanation of shaded areas, see text.} 
\label{fig:bc1}
\end{figure}

\begin{figure}
\includegraphics[width=0.46\textwidth,angle=-90]{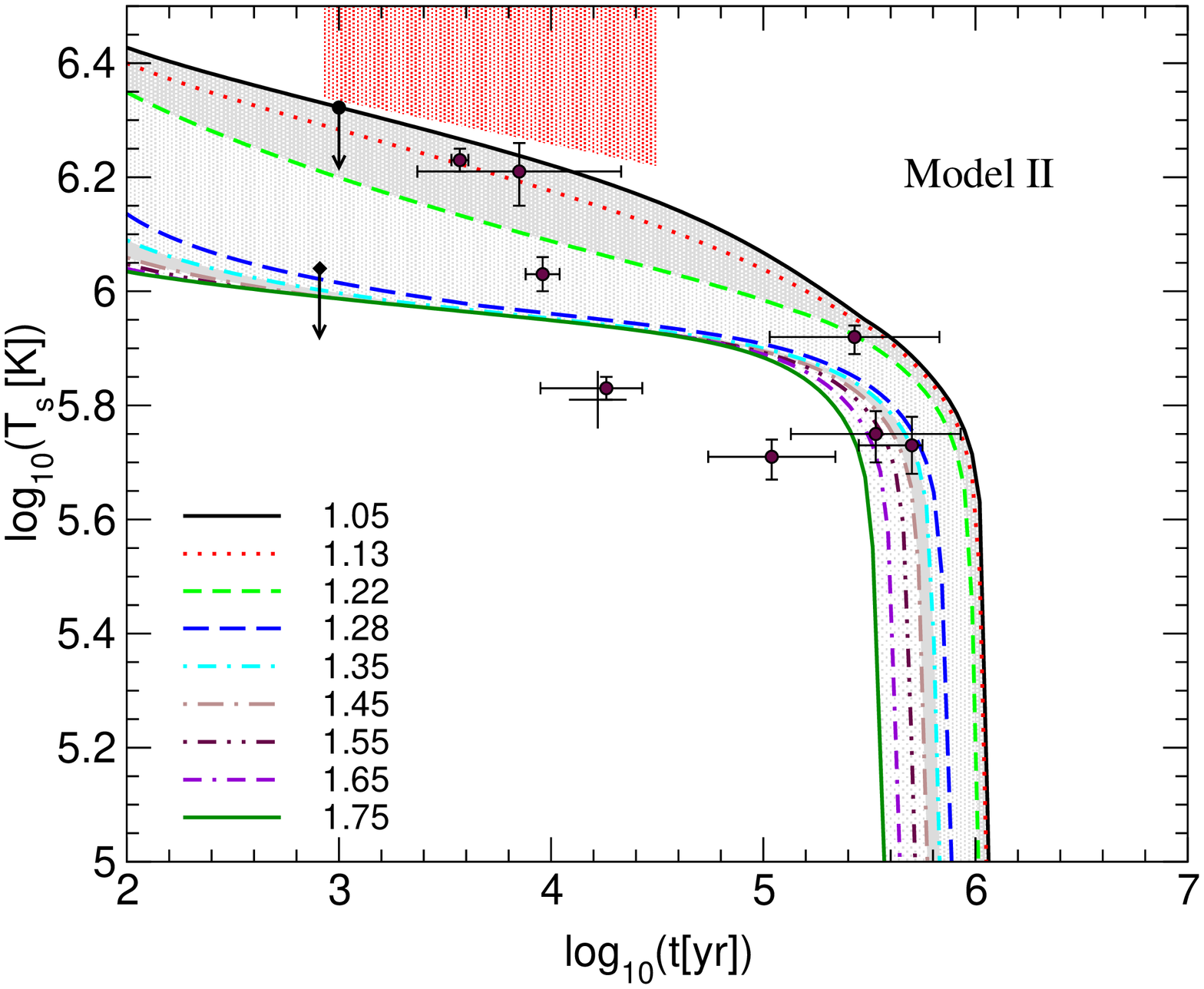}
\caption[]{Same as Fig. \ref{fig:bc1} for Model II.} \label{fig:bc2}
\end{figure}

\begin{figure}
\includegraphics[width=0.46\textwidth,angle=-90]{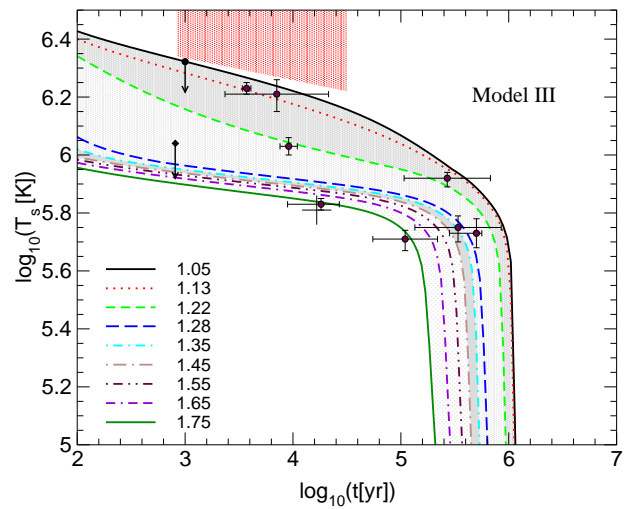}
\caption[]{Same as Fig. \ref{fig:bc1} Model III. } \label{fig:bc3}
\end{figure}

\begin{figure}
\includegraphics[width=0.46\textwidth,angle=-90]{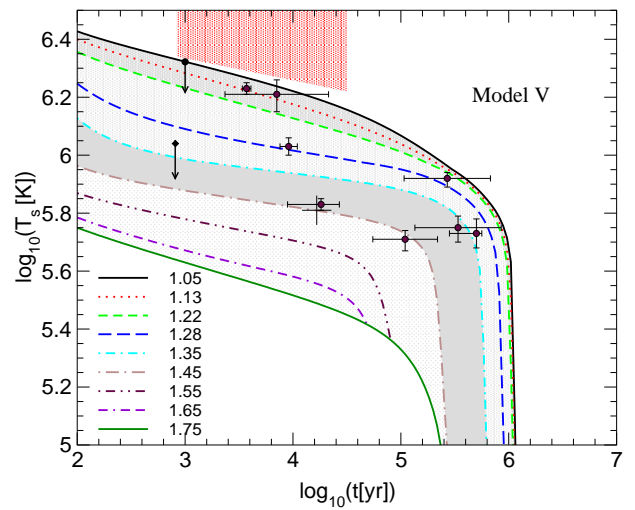}
\caption[]{Same as Fig. \ref{fig:bc1} for Model IV. } \label{fig:bc4}
\end{figure}

In the figures we present $\mathrm{T}$-$\mathrm{t}$ plots for
four models used in this paper. On each plot data points for
known cooling NSs are added (see details in \cite{Grigorian:2005fd}).
The hatched trapeze-like region represents the brightness constraint
(BC). For each model nine cooling curves are shown for configurations 
with mass values corresponding to the binning of the population synthesis 
calculations explained in the next section. 

Clearly, all models satisfy the BC. As for the
$\mathrm{T}$-$\mathrm{t}$  test the situation is different. The
Model II does not pass the test because even the highest mass configuration
(which corresponds to the coolest HyS) cannot explain the lowest
data points. So, in the Table~1 it is marked that the model does
not pass the test.

In this work we want to introduce a more detailed measure for the 
ability of a cooling model to describe observational data in the 
temperature-age diagram.
We assign five grey values to regions of compact star masses in the 
$T-t$ diagram which encode the likelihood that stars in that 
mass interval can be found in the solar neighborhood, according to the 
population synthesis scenario, see Fig. \ref{fig:mass}. 
The darkest grey value, for example, corresponds to the mass interval 
$1.35 \div 1.45$ M$_\odot$ for which the population sysnthesis predicts 
the most objects. 
According to this refined mass spectrum criterion a cooling model is 
optimal when the darkness of the grey value is in accordance with the 
number of observed objects in that region of the temperature-age diagram.
This criterion is ideally fulfilled for Model IV, where with only one exception
all objects are found in the two bands with darker grey values
whereas for Models I - III about half of the objects are situated in 
light grey or even white regions.

\section{Population synthesis scenario}
\label{pop}

Population synthesis is a frequently used technique in astrophysics 
described, e.g., in the review \cite{Popov:2004gw} where further 
references can be found.
The idea is to construct an evolutionary scenario for an artificial
population of certain astronomical objects. The comparison with observations
gives the opportunity to test our understanding of evolutionary laws and
initial conditions for these sources.

The scenario that we use in this paper is nearly identical to the one
used in Paper I. We just briefly recall the main elements and
then describe the only small difference in the mass spectrum.

The main ingredients of the population synthesis model we use are: 
the initial
distribution of NSs and their birth rate; the velocity distribution of NSs;
the mass spectrum of NSs; cooling curves and interstellar absorption. 

In this series of papers in which we use the population synthesis model as a
test of the theory of thermal evolution of NSs,
we assume that the set of cooling curves is the most
undetermined igredient. So, we make an attempt to test it. 
The cooling curves used in this paper are described in Sec. 2. 

We assume that NSs are born in the Galactic disc and in the Gould Belt.
The disc region is calculated up to 3 kpc from the Sun, and is assumed to be
of zero thickness. The birth rate in the disc part of the distribution
is taken as 250 NS per Myr. 
The Gould Belt is modeled as a flat disc-like structure with a hole
in the center (see the Belt description in \cite{p97}). 
The inclination of the Belt relative to the galactic plane is
$18^{\circ}$.  The NS birth rate in the Belt is 20 per Myr. 

The velocity distribution of NSs is not well known. 
In our calculations we use the one proposed by \cite{Arzoumanian:2001dv}. 
This is a bimodal distribution with two
maxima at $\sim 127$ and $\sim707$~km~s$^{-1}$. 
Recent results question this bimodality \cite{Hobbs:2005yx}. 
However, since the time scales in our calculations typically
are not very long, the exact form of the distribution is not very important. 

For the calculation of the column density towards a given NS we use the same
approximation as we used before. 
It depends only on the distance to the galactic center and the height above 
the galactic plane (see Fig.~1 in \cite{Popov:1999qn}). 
The detailed structure of the interstellar medium (ISM) 
is not taken into account, except the Local Bubble, which is
modeled as a 140-pc sphere centered on the Sun. 
   
The mass spectrum is a crucial ingredient of the scenario. 
The main ideology of its derivation
is the same as given in \cite{Popov:2003iq}. 
At first we take all massive stars which can produce 
a NS (spectral classes B2-08) 
from the HIPPARCOS catalogue with parallaxies $<0.002$~arcsec.
Then for each spectral class we assign a mass interval.
In the next step  using calculations by \cite{woosley} we obtain 
the baryon masses of compact remnants. 
Then we have to calculate the gravitational mass.
In this paper, unlike our previous studies where we just used the formula
from \cite{Timmes:1995kp} 
$M_{\mathrm{bar}}-M_{\mathrm{grav}}=\alpha \, M_{\mathrm{grav}}^2$ with
$\alpha=0.075$, we use $M_{\mathrm{grav}}$ accurately calculated for the
chosen configuration. 
Since the approximation from \cite{Timmes:1995kp} is 
very good the difference with the mass spectrum we used in Paper I is
tiny, and appears only in three most massive bins of our spectrum.

\begin{figure}[t]
\includegraphics[width=0.46\textwidth,angle=0]{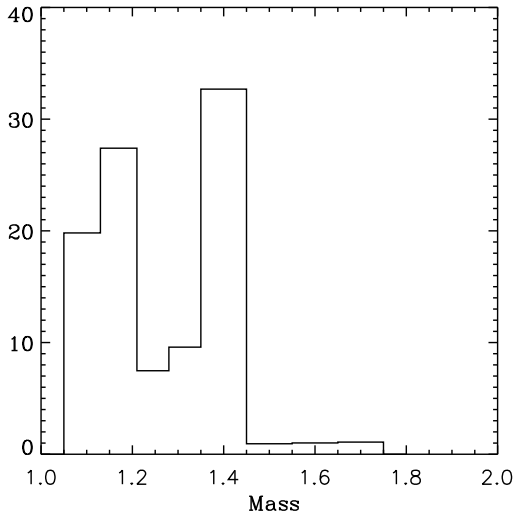}
\caption[]{The adopted mass spectrum, binned over eight intervals of
different widths. The non-truncated spectrum is shown (see text).
}
\label{fig:mass}
\end{figure}

In this paper we slightly rebinned 
the mass spectrum in order to  have a better
coverage of the cooling behaviour for the chosen configurations. 
As before we use eight mass bins defined by their borders:
1.05; 1.13; 1.22; 1.28; 1.35; 1.45; 1.55; 1.65; 1.75~$M_{\odot}$, 
see Fig.~\ref{fig:mass}. 
The critical mass for the formation of a quark core is close to 
1.22~$M_{\odot}$.
Therefore, bins are chosen such that the first two represent purely
hadronic stars. 
Bins are of different width. 
Outermost bins have a width 0.1~$M_{\odot}$. 
We do not expect HyS with masses M $<1.05$ M$_{\odot}$. 
The upper boundary of the eighth bin lies close to the
maximum mass allowed by the chosen configuration.  

Following the suggestion by \cite{Timmes:1995kp} we make runs for two 
modifications of the mass spectrum. 
Except the usage of the full range of masses we
produce, in addition, calculations for the truncated spectrum. 
In this case contributions of the first two bins are added to the third bin. 
This situation reflects the possibility that stars with 
$M\stackrel{<}{\sim} 11 \, M_{\odot}$ 
can produce NSs of similar masses close to ${\sim} 1.27\, M_{\odot}$.

As in Paper I we neglect effects of a NS atmosphere, and use pure blackbody
spectra.  As we do not address particular sources this seems to be a valid
approximation.

The population synthesis code calculates spatial trajectories of NSs
with the time step $10^4$~yrs. For each point from the set of cooling curves
we have the surface temperature of the NS. 
Calculations for an individual track are stopped at the age 
when the hottest  NSs (for all five models here this is a star with $M=1.1\,
M_{\odot}$, unless the truncated mass spectrum is used) 
reaches the temperature $10^5$~K.
Such a low temperature is beyond the registration limit for ROSAT even for 
a very short distance from the observer.   
With the known distance from the Sun and the ISM
distribution   we calculate the column density. Finally, count rates are
calculated using the ROSAT response matrix. Results are summarized along
each individual trajectory. We calculate 5,000 tracks for each model. Each
track is applied to all eight cooling curves. 
With a typical cooling timescale of about 1~Myr, we obtain $\sim 4 \cdot 10^6$ 
``sources''. 
The results are then normalized to the chosen NS formation rate 
(290 NSs in the whole region of the problem).


\section{Numerical results}
\label{res}

In this section we present results of our calculations for four models,
characterized by different sets of the two-parameter ansatz for the X-gap, 
Eq. (\ref{gap}),  see Table 1.
$\mathrm{Log\, N}$-$\mathrm{Log\, S}$ curves are given for two values
of the Gould Belt radius ($R_{\mathrm{belt}}=300$ and 500~pc), 
and for two variants of the mass spectrum (full and truncated).

The modeled $\mathrm{Log\, N}$-$\mathrm{Log\, S}$ curves are confronted with 
data for close-by, young cooling NSs observed by ROSAT.
This data set includes the {\it Magnificent Seven}  
(seven dim radio-quiet NSs),
radio pulsars, Geminga and a geminga-like source (see the list and details
in \cite{Popov:2003hq}). 
The error bars correspond to poissonian errors (i.e. square
root of the number of sources).
An important upper limit is added \cite{Rutledge:2003kg} which   
represents an estimate of unidentified cooling NSs in the ROSAT Bright
Source Catalogue (BSC). 

Model I is the best model from Paper II.
An important feature of this model is that cooling curves cover data points
in the $\mathrm{T}$-$\mathrm{t}$ plot very uniformly. 

The parameters of the Model II  were specifically chosen in such a way,
that it is possible to demonstrate the fact that the $\mathrm{Log\,
N}$-$\mathrm{Log\, S}$ test can be successful for a set that fails to pass
 the $\mathrm{T}$-$\mathrm{t}$ test. Even for the highest possible mass it
is imposible to explain cold stars, but as the $\mathrm{Log\,
N}$-$\mathrm{Log\, S}$ test is not sensitive to what is happening with
massive NSs it does not influence the results of the population
synthesis.

Model III is an attempt of a compromise to fulfill at least marginally all
three tests. It has s smaller gap than the Model I, and unlike Model II it
has a non-zero value of $\alpha$. It can explain all data points in the   
$\mathrm{T}$-$\mathrm{t}$ plot within an available mass range resorting,
however, to the very unlikely objects with masses above 1.5 M$_\odot$.

Models IV  assumes a steeper density dependence of the X-gap than all 
previous ones. It is thus possible to spread the set of cooling curves over 
the existing cooling data already for mass variations within the range of 
most probable mass values $1.25\pm 0.25$ M$_\odot$ in Fig. \ref{fig:mass}.
Using these models it is possible to describe even the Vela pulsar being a 
young, nearby and rather cool object, within this mass range.

\section{Discussion}
\label{disc}

In the present paper we were to find a set of parameters
for which all three tests ($\mathrm{T}$-$\mathrm{t}$, 
$\mathrm{Log\, N}$-$\mathrm{Log\, S}$, and BC) can be successfully passed?
Are these tests sufficient to constrain a model or is it necessary to assume 
additional constraints?

For the $\mathrm{T}$-$\mathrm{t}$ test it is necessary to cover the observed
points by curves from a relatively wide range of masses which are
consistent with known data on inital NS mass distribution (for example, data
on masses of not-accreted, i.e. mainly secondary, 
companions of double NS systems). 
For the Model III (see Fig.~\ref{fig:bc3}), 
for example, this is not the case as a number of data
points seems to correspond to a narrow mass range slightly below the critical 
mass $1.22 \, M_{\odot}$.
On the one hand, all data points on Fig.~\ref{fig:bc3}
can be covered by cooling curves from the standard 
mass range ($\sim 1$~-~$1.5, M_{\odot}$). On the other hand, the intermediate
region (${\mathrm{log}}\,T\sim 6$~-~$6.2$, 
${\mathrm{log}}\,t\sim 3$~-~$4$) corresponds to a narrow mass range: 
$\sim1.22$~-~1.26~$M_{\odot}$.
This is not a dramatic disadvantage, especially if the hypothesis discussed
in \cite{woosley,Timmes:1995kp} that stars with masses below 
$\sim 11\, M_{\odot}$
form remnants of nearly the same mass close to the range above.
Still, this property of cooling curves should be mentioned.
Model IV (Fig.~\ref{fig:bc4}) gives a more appropriate
description from the point of view of the mass distribution.

\begin{figure}
\includegraphics[width=0.46\textwidth,angle=-90]{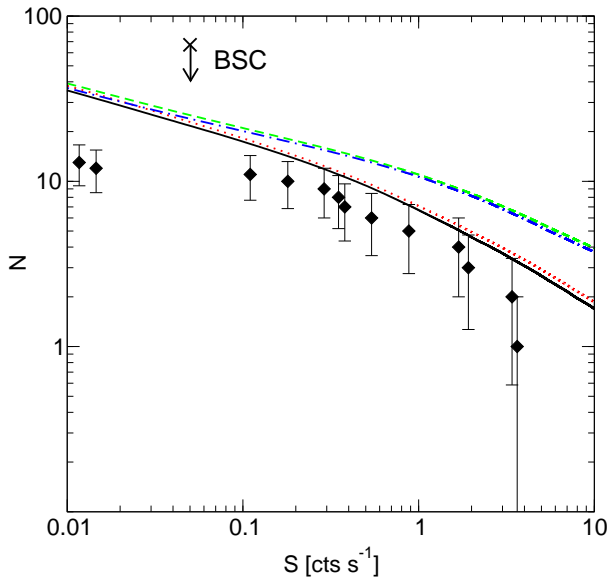}
\caption[]{ 
$\mathrm{Log\, N}$-$\mathrm{Log\, S}$ distribution for Model I.
Four variants are shown:
$\mathrm{R_{belt}}=500$~pc and truncated mass spectrum (full line),
$\mathrm{R_{belt}}=500$~pc and non-truncated mass spectrum (dotted line),
$\mathrm{R_{belt}}=300$~pc and truncated mass spectrum (dash-dotted line),
and finally $\mathrm{R_{belt}}=300$~pc (dashed line)
for non-truncated mass distribution.}
\label{fig:m1}
\end{figure}

\begin{figure}
\includegraphics[width=0.46\textwidth,angle=-90]{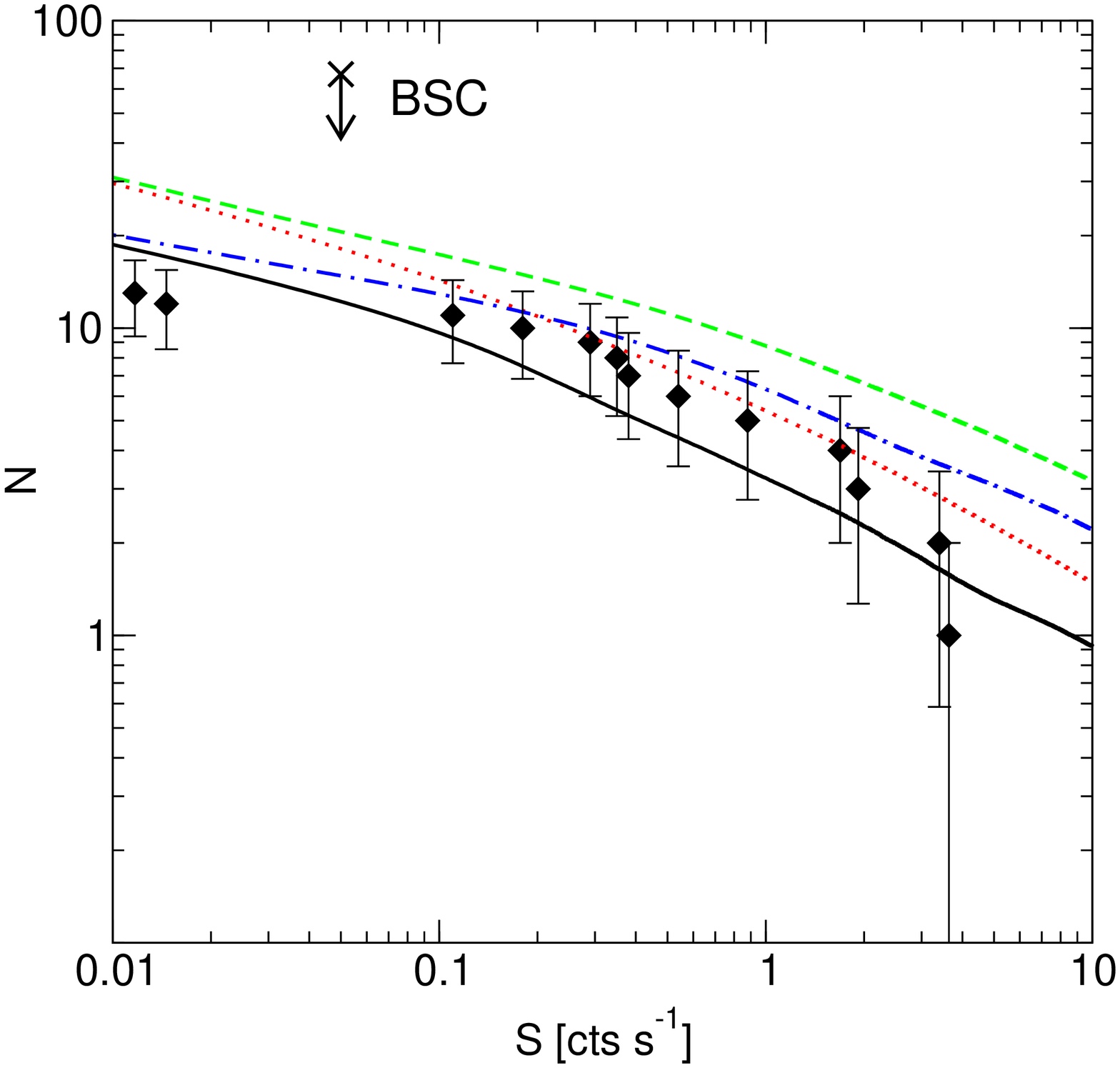}
\caption[]{
$\mathrm{Log\, N}$-$\mathrm{Log\, S}$ distribution for Model II.
Line styles as in Fig. \ref{fig:m1}.}
\label{fig:m2}
\end{figure}
%

\begin{figure}
\includegraphics[width=0.46\textwidth,angle=-90]{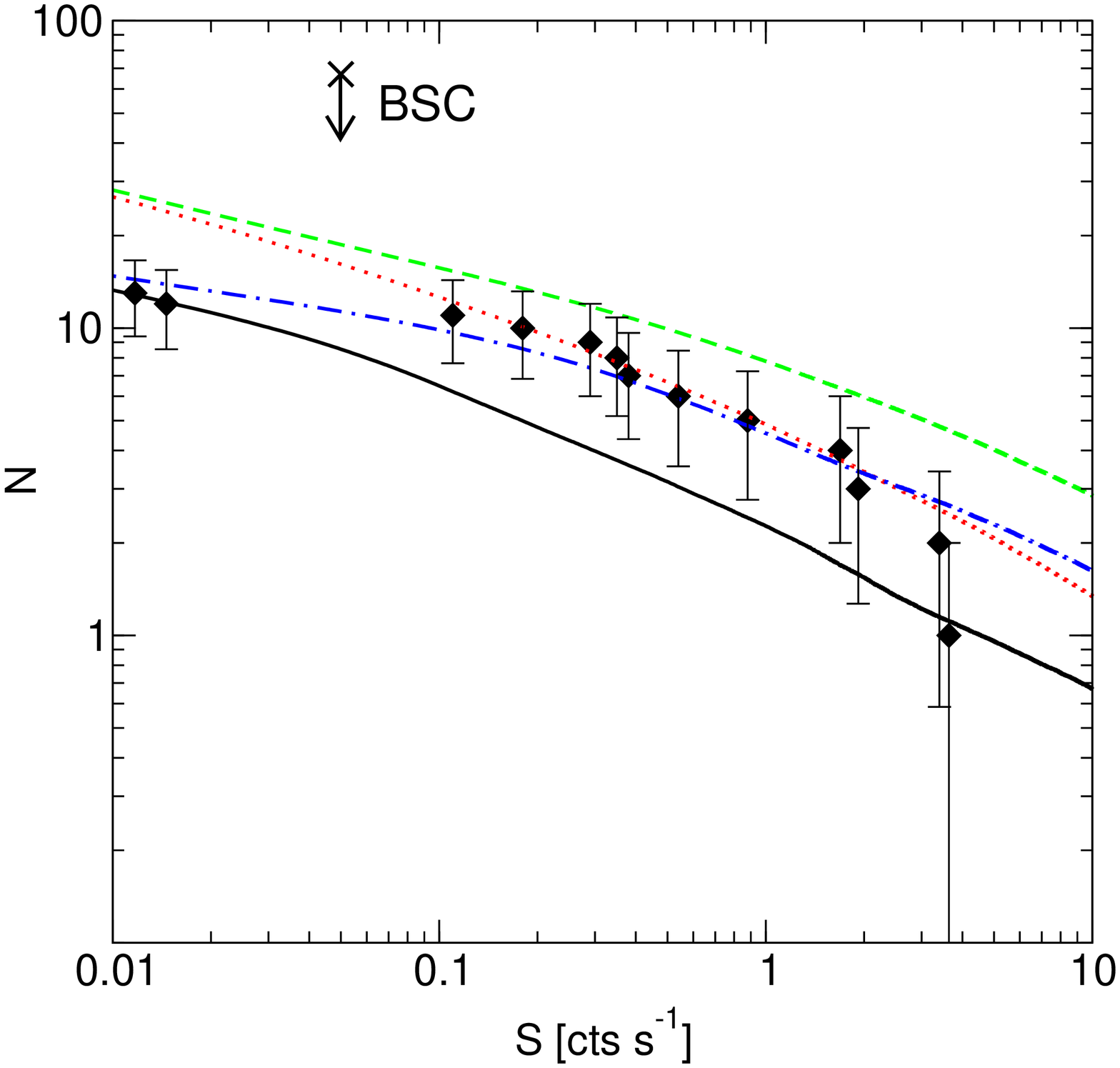}
\caption[]{$\mathrm{Log\, N}$-$\mathrm{Log\, S}$ distribution for Model III.
Line styles as in Fig. \ref{fig:m1}.}
\label{fig:m3}
\end{figure}
%
\begin{figure}
\includegraphics[width=0.46\textwidth,angle=-90]{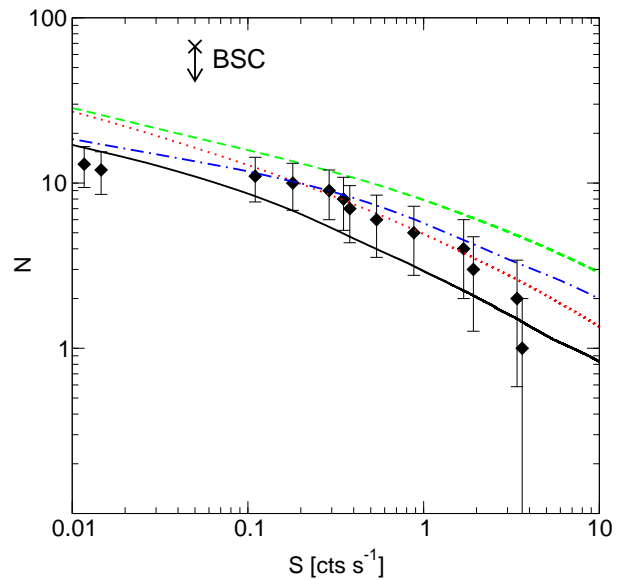}
\caption[]{$\mathrm{Log\, N}$-$\mathrm{Log\, S}$ distribution for Model IV.
Line styles as in Fig. \ref{fig:m1}.}
\label{fig:m4}
\end{figure}

For the case of  the $\mathrm{Log\, N}$-$\mathrm{Log\, S}$ test we'd like to
note that the value $\mathrm{R_{belt}}$=300~pc is more reliable. 
So, the Model I, for
which at bright fluxes we see an overprediction of sources for this value of
the Gould Belt radius, can be considered only as marginally passing the
test.   Other models do better. Especially, models III and IV. 
For them $\mathrm{Log\, N}$-$\mathrm{Log\, S}$ curves for $\mathrm{R_{belt}}=300$~pc match 
well the data points, leaving a room for few new possible identifications of
close-by cooling NSs (active search is going on by different groups
in France \cite{mo2005a,mo2005b}, in Germany \cite{bp2005}, in 
Italy \cite{Chieregato:2005es} and in the USA\cite{ag2005}). 
 
As in our  $\mathrm{Log\, N}$-$\mathrm{Log\, S}$ calculations we use a
particular mass spectrum in which there are nearly no objects with 
$M\stackrel{>}{\sim} 1.4$~-~1.5~$M_{\odot}$, we have a strong constraint 
on properties of a set of cooling curves which can satisfy all three tests. 
The position of the critical curve which devides hadronic stars from HyS is 
fixed by the chosen configuration. If curves for masses up to 1.4~$M_{\odot}$
lie too close to the critical one, then we overpredict the number of sources
on the $\mathrm{Log\, N}$-$\mathrm{Log\, S}$ plot. If oppositely, we move
curves for 1.3~-1.4~$M_{\odot}$ down (this is achieved by increasing the 
parameter $\alpha$ and thus making the density dependence of the X-gap steeper)
then a narrow range of masses becomes responsible for a wide 
region in the $\mathrm{T}$~-~$\mathrm{t}$ diagram. 

A solution could be in changing the exponential dependence in Eq. 1 to the
power-law. We plan to study this possibility in future.   

The combined usage of all three tests can put additional constraints on the 
mass spectrum of compact objects. 
For example, if we look at Fig.~\ref{fig:bc3} it is clear
that small masses (M < 1.2 M$_{\odot}$) are necessary to explain hot objects
with ages $\sim10^3$~-~$10^4$~yrs. If in such a case only a model with
a truncated mass spectrum is able to explain the $\mathrm{Log\,
N}$-$\mathrm{Log\, S}$ distribution of close-by NSs then the model is in
trouble. On other hand, if an explanation of the $\mathrm{Log\,
N}$-$\mathrm{Log\, S}$ deserves stars from low-mass bins, but cooling curves
for these stars are in conradiction with the BC, then, again,
the model has to be rejected.

In this series of our studies of the local population of cooling
compact objects we use the mass spectrum which should fit to the solar
neighbourhood enriched with stars from the mass range 8~-~15~$M_{\odot}$.
Mass spectrum of all galactic newborn NSs can be different, but not
dramatically. The number of low-mass stars 
($\sim1$~-~$1.3 \, M_{\odot}$) can be
slightly decreased in favour of more massive stars. 
However, compact objects with $M\stackrel{<}{\sim} 1.5 \, M_{\odot}$ 
anyway should  significantly outnumber more massive objects. 
This claim has some observational support.

Unfortunately, a mass determination with high precision is available only 
for NSs in binary systems.
Compact objects in X-ray binaries could accrete a significant amount of
matter. Also mass determinations for them are much less precise than for
radio pulsar systems. So, we  concentrate on the latter. 
For some of the radio pulsars observed in  binaries, accretion also played an
important r\^ole. Without any doubts masses of millisecond  pulsars do not
represent their initial values. However, there is a small number of NSs with
well determined masses, for which it is highly possible that these masses
did not change significantly since these NSs were born 
(data on NS masses can be found, for example, in 
\cite{Manchester:2004bp,cordes05,Lorimer:2005bw} and references therein).
These are secondary (younger) components of double NS systems. 
According to standard evolutionary scenarios these compact
objects never accreted a significant amount of mass (as when they formed the
primary component already was a NS). Their masses lie in the narrow range
1.18~-~1.39~$M_{\odot}$. Primary components of double NS binaries could
accrete during their lifetime. However, this amount of accreted matter
cannot be large as these are all high-mass binaries.
Masses of these NSs are all below 1.45~$M_{\odot}$. This is also an
important argument in favour of the statement that  initial masses of most 
NSs are below $\sim 1.4$~-~$1.5 \, M_{\odot}$. 

The recently discovered highly relativistic 
binary pulsar J1906+0746 \cite{Lorimer:2005un} is the nineth example of such 
a system. The total mass is determined to be 
$2.61\pm 0.02 \, M_{\odot}$.  
The pulsar itself is a young, not millisecond, object. 
It should not increase its mass due to accretion. 
So, we can assume that it is at least not heavier than the second - 
non-pulsar - component (we neglect here the possibility that the 
companion is a massive white dwarf, still this is a possibility), 
than we obtain that its mass is $\stackrel{<}{\sim} 1.3 \, M_{\odot}$. 
Nine examples (without any counter-examples) is a very good evidence in favour
of the mass spectrum used in our calculations. Of course, some effects of  
binary evolution can be important (for example, the
mass of the core of a massive star can be influenced if the star looses part 
of its mass due to mass transfer to the 
primary companion or due to common envelope formation), and so for isolated 
stars (or stars in very wide binaries) 
the situation can be slightly different. 
However, with these observational  
estimates of initial masses of NSs we feel
more confident using the spectrum with a small number of NSs with  
$M \stackrel{>}{\sim} 1.4$~-~$1.5 \, M_{\odot}$. 

Brighter sources are easier to discover. So, among known cooling NSs the
fraction of NSs with masses 
$1 \, M_{\odot}\stackrel{<}{\sim} M \stackrel{<}{\sim} 1.5 \, M_{\odot}$ 
should be even higher than in the original mass spectrum. 
So, we have the impression that it is
necessary to try to explain even cold (may be with an exception of 1-2
coldest) sources with $M\stackrel{<}{\sim}  1.4$~-~$1.5 \, M_{\odot}$.
Especially, the Magnificent seven and other young close-by 
compact objects should be
explained as most typical representatives of the whole NS 
population \cite{footnote}.
We want to underline that, even being selected by their observability in
soft X-rays, these sources form one of the most uniform samples of young
isolated NSs.  
In this sense, the situation as in Fig.~\ref{fig:bc1} where a significant
number of sources are explained by cooling curves corresponding to 
$1.5 \, M_{\odot} \stackrel{<}{\sim} M \stackrel{<}{\sim} 1.7\, M_{\odot}$ 
should be considered as a disadvantage of the model. 
Particularly, Vela, Geminga and RX J1856-3754 should not be
explained as massive NSs. These all are young close-by sources, and the 
probability that so near-by we observe young NSs which come out of few
percent of the most massive objects of this class is low.
     
All the above gives us the opportunity to formulate the conjecture of a
{\it mass spectrum constraint}: data points should be explained mostly by 
NSs with {\it typical} masses. 
For all known data this {\it typical} means 
$1.1 \, M_{\odot} \stackrel{<}{\sim} M \stackrel{<}{\sim}  1.5 \, M_{\odot}$.

 The $\mathrm{Log\, N}$-$\mathrm{Log\, S}$ test is the only one that takes
into account the mass spectrum {\it explicitely}.  This is an additional
argument  in favour of using this test together with others.

 Taking all together, we conclude that Model IV is the best among studied
examples from the point of view of all three tests plus mass constraint.

\section{Conclusion}

We made a preliminary study of cooling curves for HyS based on the approach
in Paper II which suggests that if quark matter occurs in a compact star,
it has to be in the 2SC+X phase, where the hypothetical X-gap still lacks a 
microscopical explanation. 
All three tests of the cooling behavior 
($\mathrm{Log\, N}$-$\mathrm{Log\, S}$, 
$\mathrm{T}$-$\mathrm{t}$, BC) are applied.
Four models defined by the two-parameter ansatz for the X-gap were calculated. 
Model II with a density-independent X-gap could directly be excluded since it 
was not able to explain some cooling data including Vela at all.
Two of the models (I and III) successfully passed two tests and marginally the 
third. None of these models could explain explain the temperature-age value 
for Vela within a typical mass range, i.e. for a Vela mass below 1.5 M$_\odot$.
However, with a steeper density dependence of the X-gap than suggested in 
Paper II, we were able to fulfill all 4 constraints and 
exemplified this for model IV.
To conclude, HyS with a 2SC+X quark matter core appear to be good candidates  
to explain the cooling behaviour of compact objects, although a consistent 
theoretical explanation of the 
hypothetical X-gap and its steep density dependence has still to be developed. 

\begin{acknowledgements}
We thank M.E. Prokhorov, R. Turolla, D.N. Voskresensky and F. Weber 
for their discussions and contributions to this research field. 
S.P. thanks for the hospitality extended to him during a visit at
the University of Rostock, and for support from the DAAD partnership program
with the Moscow State University. His work was supported in part by RFBR 
grants No. 04-02-16720, 06-02-16025 and by the ``Dynasty'' foundation.
The work of H.G. was supported by DFG grant No. 436 ARM 17/4/05 and he 
acknowledges hospitality and support by the Bogoliubov Laboratory for 
Theoretical Physics at JINR Dubna where this work has been completed.

\end{acknowledgements}

\begin{table*}[h]
\begin{tabular}{|l||c|c||c|c|c|c|c|}
\hline
Model& $\Delta_0$ [MeV] & $\alpha$ & BC &T -- t & Log N -- Log S & M$_{\rm typ} \le$ 1.5 M$_\odot$ & All tests\\
\hline \hline
I   &  1   & 10  & +& +   & $\circ$  & -& -\\
II  &  0.1 & 0 & +& -   & + & - & -\\
III &  0.1 &  2  & +& $\circ$ & + & -& -  \\
IV &  5 &  25  & +& + & + & + & +\\
\hline
\end{tabular}
\caption{Parameter values $\Delta_0$  and $\alpha$ for Models I - IV defining 
the density dependence of the X-gap introduced in Eq. (\ref{gap}).
The symbols +, - and $\circ$ denote whether or not a given model passes the proposed cooling tests, fails or is indifferent, respectively.
}
\end{table*}
\vspace{5mm}

\end{document}